\documentclass{article}

\usepackage{amsthm}
\usepackage{amsmath}
\usepackage{natbib}
\usepackage[colorlinks,citecolor=blue,urlcolor=blue,filecolor=blue,backref=page]{hyperref}
\usepackage{graphicx}
\usepackage{authblk}

\begin{document}

\title{Bayesian modeling of population variance for aggregated measurements.}
\date{    }
\author[1]{Elena Moltchanova (mailto: elena.moltchanova@canterbury.ac.nz)}

\author[1]{\\Daniel Gerhard}
\author[2]{Rory Ellis}

\affil[1]{School of Mathematics \& Statistics, University of Canterbury, Private Bag 4800, Christchurch 8140, New Zealand}
\affil[2]{Fonterra, Palmerston North, New Zealand}


\maketitle

\begin{abstract}
Growth curves are commonly used in modeling aimed at crop yield prediction. Fitting such curves often depends on availability of detailed observations, such as individual grape bunch weight or individual apple weight. However, in practice, aggregated weights (such as a bucket of grape bunches or apples) are available instead. While treating such bucket averages as if they were individual observations is tempting, it may introduce bias particularly with respect to population variance. In this paper we provide an elegant solution which enables estimation of individual weights using Dirichlet priors within Bayesian inferential framework.

\end{abstract}

\section{Introduction}

Crop yield prediction is important. It enables the growers to make decision about crop management throughout the season as well as to allocate resources for the final harvesting and to make decisions about eventual processing and sales. Entering the string "crop yield prediction" into Google Scholar results in over 900 publications for the year 2020 alone. Approximately one tenth of these mention grapes. Other often studied crops include apples (106 mentions), strawberries (50 mentions), peaches (31 mentions), and kiwifruit (13 mentions). The fruit growth is usually modeled using some variety of a non-linear curve (with \cite{fernandes2017double}, \cite{lakso1995expolinear}, and \cite{coombe2000dynamics} being just a few examples). For grapes, a double-sigmoidal curve \cite{coombe2000dynamics} is a common way to model both the berry and the bunch growth.

The models such as the ones in \citet{de2015comparison}, and more recently \citep{ellis2020using}, are based on grape bunch/cluster weight data collected throughout the season. Although there is an increased push towards using sensors and computer vision in identifying fruit sizes and weight (see \cite{bulanon2020machine} for a recent review), many models are still based on physical measurements of grape bunch or cluster weights.  However, in many instances, measuring each individual grape bunch (or an apple or any other fruit) becomes too arduous and expensive. Therefore in practice, bucket weights are obtained instead of individual weights. Thus, for example, rather than be provided with 20 individual bunch weights, we can instead be informed that the aggregate weight of $20$ bunches was $3000$ g. 

A quick-and-dirty solution is to simply obtain the average weight, in this case, $3000/20 = 150 g$, and treat is as a single observation. However, such simplification completely disregards the uncertainty associated with each average obtained in such a way. Although such aggregation may not impact the estimate of the mean bunch weight, it does affect the estimation of variance, and thus the perceived accuracy of the prediction, which is often the ultimate goal of the exercise. Moreover, in more complex modeling, the parameter estimates may in fact be biased as well especially if the bucket sizes differ wildly. 

Therefore, in this manuscript, we propose an elegant solution to this problem. We use a Bayesian framework to assign a Dirichlet prior to the distribution of individual bunch weights within the bucket, and construct an MCMC algorithm to estimate them. We then use simulation studies to illustrate the performance of the proposed method.

\section{Theory}

Consider a continuous random variable

\begin{equation*}
x \sim N(\mu,\tau),
\end{equation*}

where $\mu$ is the mean and $\tau$ is the precision (i.e. inverse variance) of the normal distribution. An average of a sample of size $n$, $\bar{x}_n$ will then also have a normal distribution:

\begin{equation*}
\bar{x} \sim N(\mu,n\tau).
\end{equation*}

Assume a conjugate normal prior for the normal mean $\mu$:

\begin{equation*}
\mu \sim N(\mu_0,\tau_0),
\end{equation*}

and a conjugate Gamma prior for the normal precision $\tau$:

\begin{equation*}
\tau \sim Gamma(a,b).
\end{equation*}

If our data consist of $K$ samples of sizes $n_1, ...,n_K$ with the respective averages $\bar{x}_1,...,\bar{x}_K$, we can derive

\begin{equation*}
\bar{x}_k \sim N(\mu,n_k\tau),
\end{equation*}

and use Bayes' formula to derive the conditional posterior distributions as:

\begin{equation}
\mu|\cdot \sim N\left(\frac{\tau \sum_kn_k\bar{x}_k+\tau_0\mu_0}{\tau \sum_kn_k+\tau_0},\tau \sum_kn_k+\tau_0\right)
\end{equation}

and

\begin{equation}
\tau|\cdot \sim Gamma\left(a+0.5K,b+0.5\sum_kn_k(\bar{x}_k-\mu)^2\right).
\end{equation}

These conditional distributions can then be used in the context of Gibbs sampling to produce the posterior distributions for the parameters $\mu$ and $\tau$.

However, it is not always possible to derive analytically the distribution of a sum (or a sample average) of observations.

Consider the following model, where the observations have some parametric p.d.f. $g(x_{ij}|\theta)$ with $j=1,...,n_i$ and $i=1,...,K$ dependent on parameter $\theta$. Let the observations be aggregated into $K$ groups of perhaps different sizes $n_1, n_2,..., n_K$ where only the sums 

\begin{equation*}
y_i = \sum_{j=1}^{n_i}x_{ij} \qquad j=1,...,n_i, \textrm{ and }  i=1,...,K,
\end{equation*}

are reported. 

Given the prior distribution $p(\theta)$, within the Bayesian framework, one can construct an MCMC algorithm to produce a sample from the posterior distribution 

\begin{equation} \label{eq:MCMCposterior}
p(\theta|\mathbf{x}) \propto \prod_i\prod_jg(x_{ij}|\theta)p(\theta).
\end{equation}

When $x_{ij}$ are not observed directly, one can attempt to derive the distribution of the aggregate $y_i$ and substitute it into Equation \ref{eq:MCMCposterior}. However, since this is not always possible, we propose to amend the Metropolis-Hastings algorithm to recover the $x_{ij}$ observations. In this case, the latent observations $x_{ij}$ are treated as parameters.

To make this explicit, we add the following equation to our model:

\begin{equation}
p(y|x) = Pr(y_i=y'|x_{i1},...,x_{in_i}) = \begin{cases}
    1       & \quad  \textrm{if } \sum_{j=1}^{n_i}x_{ij} = y',\\
    0  & \quad \text{otherwise.} 
 \end{cases}
\end{equation}

for $i=1,...,K$.

We can thus write out the joint posterior distribution for the model as

\begin{equation*}
p(\theta,\mathbf{x}|\mathbf{y}) \propto p(\mathbf{y}|\mathbf{x})p(\mathbf{x}|\theta)p(\theta).
\end{equation*}

Now, for a Metropolis-Hastings step, we can propose new values of $\mathbf{x^{\ast}}$ given the current ones, using a proposal distribution $q(\mathbf{x^{\ast}}|\mathbf{x})$, and evaluate the rejection ratio as:

\begin{eqnarray}
\label{eq:Rratio}
R &=& \frac{p(\mathbf{y}|\mathbf{x^{\ast}})p(\mathbf{x^{\ast}}|\theta)p(\theta)}{p(\mathbf{y}|\mathbf{x})p(\mathbf{x}|\theta)p(\theta)}\frac{q(\mathbf{x}|\mathbf{x^{\ast}})}{q(\mathbf{x^{\ast}}|\mathbf{x})}\nonumber\\
&=& \frac{p(\mathbf{x^{\ast}}|\theta)}{p(\mathbf{x}|\theta)}\frac{q(\mathbf{x}|\mathbf{x^{\ast}})}{q(\mathbf{x^{\ast}}|\mathbf{x})}
\end{eqnarray}

Note, that if $\sum_{j=1}^{n_i}x^{\ast}_{ij} \neq y_i$ for all $i=1,...,K$, then $R=0$ and the proposal will be rejected. Thus, intuitively, a good proposal distribution would ensure the correct sums, while sampling around the current values of $x_{ij}$. We thus propose the following sampling scheme.

\begin{enumerate}
\item Sample the weights $w_{ij}$ from a Dirichlet distribution
\begin{equation}
w_{i1}, ..., w_{in_i} \sim \text{Dirichlet} (\delta_i x_{i1},...,\delta_i x_{in_i}), \qquad i=1,...,K,
\end{equation}
for some chosen $\delta_1,...,\delta_K$, and $\sum_{j=1}^{n_i} w_{ij} = 1$. 
\item Evaluate the proposed values as
\begin{equation*}
x^{\ast}_{ij} = w_{ij}y_i \qquad \forall i,j.
\end{equation*}
\end{enumerate}

Note, that since the weights for each group $i=1,...,K$ come from a Dirichlet distribution

\begin{equation*}
\sum_{j=1}^{n_i}x^{\ast}_{ij} = \sum_{j=1}^{n_i}w_{ij}y_i = y_i.
\end{equation*}

The expected value of each proposed element is the current value:

\begin{equation*}
E(x^{\ast}_{ij}|x_{ij}) = y_i\frac{\delta x_{ij}}{\sum_{j=1}^{n_i}\left(\delta x_{ij}\right)} = y_i \frac{x_{ij}}{y_i} = x_{ij},
\end{equation*}
and the variance
\begin{equation*}
Var(x^{\ast}_{ij}|\mathbf{x}) = y_i^2\frac{\frac{\delta x_{ij}}{\delta y_i}\left(1-\frac{\delta x_{ij}}{\delta y_i}\right)}{\delta y_i+1}=\frac{ x_{ij}( y_i- x_{ij})}{\delta y_i+1}
\end{equation*}
increases when $\delta$ decreases.

The density function for the proposal distribution for one group $q(\mathbf{x_i^\ast}|\mathbf{x_i})$ can be written as

\begin{equation*}
q(\mathbf{x^{\ast}_i}|\mathbf{x_i}) = \frac{\Gamma(\delta y_i)}{\prod_{j=1}^{n_i}\Gamma(\delta x_{ij})}\prod_{j=1}^{n_i}(x^{\ast}_{ij})^{\delta x_{ij}-1},
\end{equation*}

where $\Gamma()$ is the gamma function. By substituting the relevant expressions into Equation \ref{eq:Rratio} for the rejection rate, and accepting the proposed values of $\mathbf{x}$ with probability $\min(R,1)$, one can estimate these latent variables.

The overall estimation algorithm consists of two parts: (i) sample $\mathbf{x}$ given $\mathbf{y}$ and $\theta$, and (ii) sample $\theta$ given $\mathbf{x}$.

The above method is applicable whatever the likelihood $p(x|\theta)$ is.

The choice of parameters $\delta_i$ for the proposal distribution will intuitively have an effect on the convergence and acceptance ratio of the algorithm. Note, that it does not have to be the same for all groups $i=1,...,K$. However, if the groups are of similar sizes, there is no reason to fine-tune group-specific $\delta_i$.

\section{Simulations. Log-normally distributed response.}

In order to demonstrate the performance of the suggested method in recovering the original population variance from the aggregate data, we have run a simulation case study for the log-normal model. Due to the computational intensity of the algorithm with the Dirichlet step, only one simulation was run for the following set-up.

$$
\log(x_i) \sim N(\mu,\tau),
$$

where $\mu = \log(250)$ and $\tau^{-\frac{1}{2}} = 0.10$, i.e., the individual bunches are expected to be within $20\%$ of the average bunch weight of $250 g$. The data were simulated for $1000$ bunches aggregated into $100$ buckets of $10$ bunches each. We have used a vague $Gamma(0.01,0.01)$ and informative $Gamma(2.5,0.025)$ priors for the precision $\tau$, and the initial values based on the actual precision and the estimate based on the aggregated sample respectively.

\begin{figure}[h!]
\centering
\includegraphics[width=0.8\textwidth]{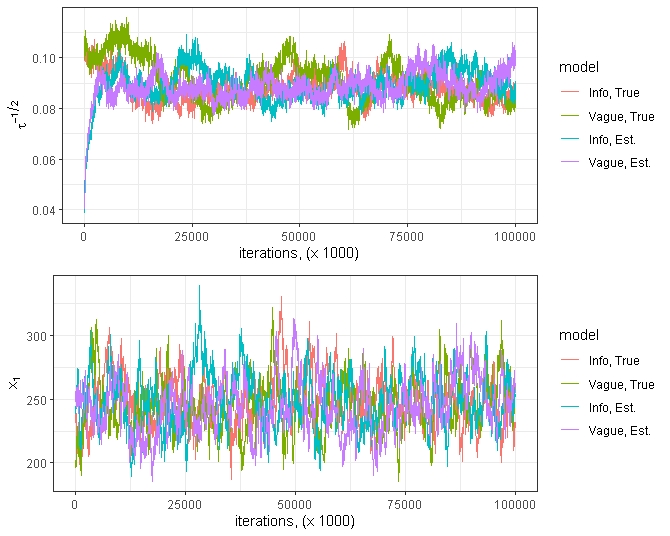}
\caption{Log-normal Model Performance: Posterior sample traces for the standard deviation $\tau^{-1/2}$ and an individual bunch $x_1$ for a single simulated data set for informative and vague priors with the starting point for $\tau$ being either the true value or the estimate based on the aggregated data..}
\label{fig:log_mix}
\end{figure}

\begin{figure}[h!]
\centering
\includegraphics[width=0.8\textwidth]{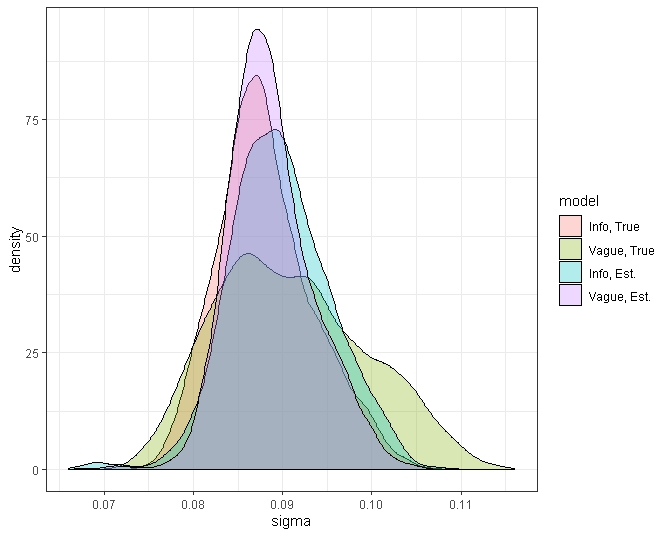}
\caption{Log-normal Model Performance: Estimated posterior distributions for the mean $\mu$ for a single simulated data set for informative and vague priors with the starting point for $\tau$ being either the true value or the estimate based on the aggregated data.}
\label{fig:performance}
\end{figure}

\begin{figure}[h!]
\centering
\includegraphics[width=0.8\textwidth]{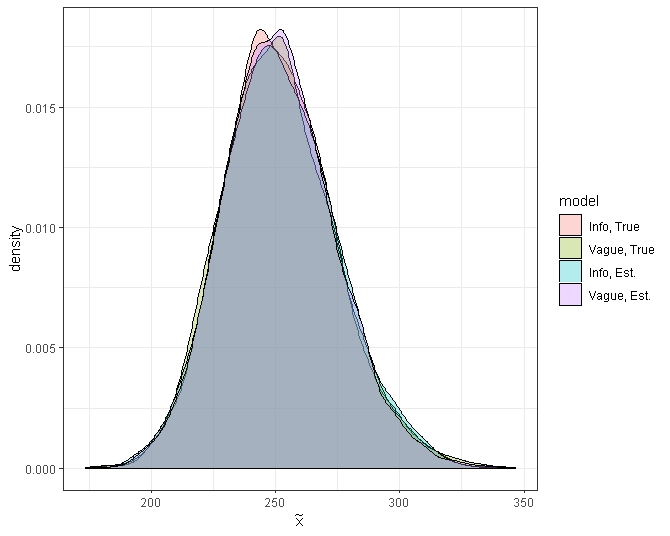}
\caption{Log-normal Model Performance: Estimated posterior predictive distributions for a random individual observation $\tilde{x}$ for a single simulated data set for informative and vague priors with the starting point for $\tau$ being either the true value or the estimate based on the aggregated data.}
\label{fig:postpred}
\end{figure}

In each case $10^8$ iterations were ran with the resulting traces shown in Figure \ref{fig:log_mix} and the posterior distributions after the burn in of $5\times10^6$ thinned at the rate of $2times10^4$ are shown in Figure \ref{fig:performance}. The estimated posterior predictive distributions for an individual observation $p(\tilde{x}|x)$ are shown in Figure \ref{fig:postpred}.

The results show that the initial values have effect on convergence. Because there is a relatively large amount of data involved in this example, the prior does not have an appreciable effect on the final result. The standard deviation $\sigma$ was accurately recovered, as was the distribution of an individual observation $\tilde{x}$.

\section{Discussion}

In this paper we have demonstrated the Bayesian method of accurately estimating the population variance based on aggregated data. While the work was motivated by an application to viticulture, it can be used more widely in any context where aggregates are recorded instead of individual observations.

Treating aggregate-based averages as if they were individual observations will clearly underestimate the underlying population variance as well as potentially bias the estimation of the population mean. It is thus clearly an easy but erroneous way to go. As we demonstrate, in some situations, such as a normally distributed response, the posterior distributions for the parameters of interest can be obtained analytically. However, in most situations, analytic derivation is not an option. The method we suggest provides an elegant solution to the problem by recovering the individual weights. 

Although the method works in a sense of recovering the original parameters, there are issues that still need to be addressed. For the log-normal model, we have found the convergence to be extremely slow, both in terms of the number iterations required and in terms of computing time need to complete those iterations. The two are obviously related, but sometimes a great number of iterations can be completed within a reasonable period of time. Unfortunately, that was not the case here and it was also the reason why our simulation study of log-normal response was very limited.

One obvious solution is to find a more efficient way to code the algorithm. Another is to make the algorithm itself more efficient. The aspect which is the most relevant to the efficient exploration of the parameter space is the proposal distribution for the Dirichlet weights, expressed via the parameter $\sigma$. It should be noted, that with each new proposal, new values for every single unobserved individual bunch weight is provided. In our simulation study, we've had total of $1000$ bunches. However, in practice, it can easily be many more. Sampling so many parameters simultaneously often means that the deviations from the status quo need to be tiny for the proposal to be accepted; especially so one the sampler has converged. However, it may also result in slow mixing and a very long burn-in, particularly if the starting values are far away from the truth. This need to have larger steps during the burn-in phase and shorter steps during the convergence phase may be accommodated via an adaptive MCMC algorithm (\cite{andrieu2008tutorial}, \cite{roberts2009examples}), and will be the next obvious step in this work.

Despite the large space for improvement in terms of computational efficiency, we believe that the proposed method demonstrates a nice way to solve the problem of incorporating the aggregated measurements into the model. Its immediate practical applications include extending the double sigmoidal grape bunch growth model of \cite{ellis2020using}, but we are certain it will find use in other agricultural applications and beyond.

\clearpage

\bibliographystyle{apalike}
\bibliography{references}

\begin{thebibliography}{}

\bibitem[Andrieu and Thoms, 2008]{andrieu2008tutorial}
Andrieu, C. and Thoms, J. (2008).
\newblock A tutorial on adaptive {MCMC}.
\newblock {\em Statistics and computing}, 18(4):343--373.

\bibitem[Bulanon et~al., 2020]{bulanon2020machine}
Bulanon, D.~M., Hestand, T., Nogales, C., Allen, B., and Colwell, J. (2020).
\newblock Machine vision system for orchard management.
\newblock In {\em Machine Vision and Navigation}, pages 197--240. Springer.

\bibitem[Coombe and McCarthy, 2000]{coombe2000dynamics}
Coombe, B.~G. and McCarthy, M. (2000).
\newblock Dynamics of grape berry growth and physiology of ripening.
\newblock {\em Australian journal of grape and wine research}, 6(2):131--135.

\bibitem[de~la Fuente et~al., 2015]{de2015comparison}
de~la Fuente, M., Linares, R., Baeza, P., Miranda, C., and Lissarrague, J.~R.
  (2015).
\newblock Comparison of different methods of grapevine yield prediction in the
  time window between fruitset and veraison.
\newblock {\em OENO One}, 49(1):27--35.

\bibitem[Ellis et~al., 2020]{ellis2020using}
Ellis, R., Moltchanova, E., Gerhard, D., Trought, M., and Yang, L. (2020).
\newblock Using bayesian growth models to predict grape yield.
\newblock {\em OENO One}, 54(3):443--453.

\bibitem[Fernandes et~al., 2017]{fernandes2017double}
Fernandes, T.~J., Pereira, A.~A., and Muniz, J.~A. (2017).
\newblock Double sigmoidal models describing the growth of coffee berries.
\newblock {\em Ci{\^e}ncia Rural}, 47.

\bibitem[Lakso et~al., 1995]{lakso1995expolinear}
Lakso, A., Corelli~Grappadelli, L., Barnard, J., and Goffinet, M. (1995).
\newblock An expolinear model of the growth pattern of the apple fruit.
\newblock {\em Journal of Horticultural Science}, 70(3):389--394.

\bibitem[Roberts and Rosenthal, 2009]{roberts2009examples}
Roberts, G.~O. and Rosenthal, J.~S. (2009).
\newblock Examples of adaptive mcmc.
\newblock {\em Journal of computational and graphical statistics},
  18(2):349--367.

\end{thebibliography}
\end{document}